\documentstyle[aps,multicol,epsfig]{revtex}  
  
\begin{document}    
\title{Structure function of passive scalars in two-dimensional turbulence}  
\author{\sc Bruno Eckhardt and J\"org Schumacher}  
\address{Fachbereich Physik, Philipps-Universit\"at Marburg,  
D-35032 Marburg, Germany} 
\date{November 1999} 
\maketitle

\begin{abstract}  
The structure function of a scalar $\theta({\bf x},t)$, passively advected in a 
two-dimensional turbulent flow ${\bf u}({\bf x},t)$, is discussed by means of 
the fractal dimension $\delta^{(1)}_g$ of the passive scalar graph.  A relation 
between $\delta^{(1)}_g$, the scaling exponent $\zeta_1^{(\theta)}$ of the 
scalar structure function $D_1^{(\theta)}(r)$, and the structure function 
$D_2(r)$ of the underlying flow field is derived.  Different from the 3-d case, 
the 2-d structure function also depends on an additional parameter, 
characteristic of the driving of the passive scalar.  In the enstrophy inertial 
subrange a mean field approximation for the velocity structure function gives a 
scaling of the passive scalar graph with $\delta^{(1)}_g<2$ for intermediate and 
large values of the Prandtl number $Pr$.  In the energy inertial subrange a 
model for the energy spectrum and thus $D_2(r)$ gives a passive scalar graph 
scaling with exponent $\delta^{(1)}_g=\frac{5}{3}$.  Finally, we discuss an 
application to recent observations of scalar dispersion in non-universal 2-d 
flows. 
\end{abstract}  
 
\begin{multicols}{2}
\section{Introduction}   
The dynamics of a scalar field $\theta({\bf x},t)$ advected in a turbulent 
velocity field ${\bf u}({\bf x},t)$ is of practical relevance in many fields of 
current research such as air pollution or chemical reactions in the stratosphere 
in connection with the ozone hole \cite{Edo96}.  Especially for problems 
in atmospheric physics, models of two-dimensional turbulent flows give a good 
approximation of the dynamical processes and are frequently 
used\cite{Lil89,Les87}.  More recently, two-dimensional turbulence has become 
experimentally accessible in mercury layers\cite{Som86}, thin salt water 
layers\cite{Tab91,JerTab97,Goll97,Car96}, and soap 
films\cite{GhaDer89,Mar98,Riv98,Rut98}.  Two-dimensional turbulence is also 
interesting because of its fundamentally different behavior compared to the 
three-dimensional case.  Since the enstrophy is a second inviscid invariant 
beside the energy two cascades develop:  starting from a fixed, intermediate 
injection scale, energy is transported to larger spatial scales in an inverse 
energy cascade and to smaller ones in an enstrophy cascade\cite{Kra67,Bat69}. 
 
The scaling behavior of a passive scalar in a turbulent fluid was analyzed 
mainly in three dimensions where three different regimes could be identified. 
Depending on the Reynolds number of the underlying fluid turbulence and the 
ratio of the kinematic viscosity to the scalar diffusivity one distinguishes the 
viscous-convective Batchelor regime\cite{Bat59}, the inertial-convective 
regime\cite{Obu49,Cor51}, and the inertial-diffusive regime.  In 2-d the 
situation is more complicated, since already the velocity field shows a variety 
of scaling regimes.  In particular, the inverse cascade process gives rise to 
the formation of large scale vortices that change on very slow time scales only 
\cite{Ben87} and can dominate the  
dynamics of the passive scalar, at least on intermediate time scales 
\cite{Bab87,Bas94}. The formation of coherent vortices can be suppressed 
by a large scale dissipation mechanism. If this additional dissipation 
is present a statistically stationary homogeneous and isotropic 
turbulent flow field develops,  
that can be characterized by its structure function. 
We assume that a  
passive scalar in such a flow field also develops a statistically 
stationary state which can be characterized by its own structure function.   
 
The approach used to analyse the structure function of the passive scalar is 
geometric measure theory \cite{Con91,ConPro93,ConPro94,ProCon93}.  This powerful 
method allows to connect the structure function of the passive scalar to that of 
the underlying flow field and thus to link the statistical behavior of both. 
The result are scale resolved bounds on the scaling behavior.  Upper bounds are 
easiest to derive and often give very good results, see e.g.  the favorable 
comparison between theory and numerical simulations in \cite{GroLoh94}.  The 
derivation of lower bounds is possible\cite{ConPro94} but much more difficult 
and will not be attempted here.  So assuming the reliability of the upper bounds 
we would like to see how the different regimes in ${\bf u}$ are reflected in the 
scaling properties of the scalar field passively advected by the flow.  Some 
aspects of the 2-d case have been discussed previously \cite{ProCon93}, see 
below.  In addition, we would like to compare the predictions to the results of 
experiments of Cardoso {\it et al.}  \cite{Car96}, where certain discrepancies 
to theory were noted.  As we will see the discrepancies can be accounted for if 
the experimentally measured structure function is substituted for the velocity 
field. 
 
The model we consider is that of a scalar field $\theta({\bf x},t)$ 
transported in the turbulent flow field ${\bf u}({\bf x},t)$ according to 
\begin{eqnarray}  
\label{sceq}  
\frac{\partial\theta}{\partial t} + ({\bf u}\cdot{\bf \nabla})\theta  
=\kappa{\bf\nabla}^2\theta + f_{\theta}\;.  
\end{eqnarray}  
$\kappa$ denotes the diffusivity.  The force density $f_{\theta}$ models 
external boundary conditions and the driving and assures a statistically 
stationary field $\theta({\bf x},t)$.  The scalar $\theta$ is assumed to be {\it 
passive}, i.e.  it does not affect the dynamics and the statistical properties 
of the velocity field.  We assume that in the presence 
of a large scale dissipation mechanism a  
homogeneous, isotropic, and stationary 
turbulent state develops. 
The ratio of the kinematic viscosity $\nu$ to the scalar  
diffusivity 
$\kappa$ defines the Prandtl number $Pr=\nu/\kappa$ (this is the  
nomenclature used when $\theta$ is a temperature field; if it describes a 
concentration then the corresponding ratio is known as the Schmidt number). 
The scaling exponents $\zeta^{(\theta)}_n$ of the $n$-th order scalar   
structure functions, defined as  
\begin{eqnarray}  
\label{strucdef}  
D^{(\theta)}_{n}(r)=\langle|\theta({\bf x}+{\bf r},t)-\theta({\bf  
x},t)|^n\rangle \sim r^{\zeta^{(\theta)}_n}\;,  
\end{eqnarray}  
can be obtained from an analysis of the fractal dimension $\delta_g^{(d)}$ of 
$d$-dimensional scalar field graphs; $\langle\cdot\rangle$ denotes the 
statistical ensemble average.  The fundamentals of the geometric measure theory 
approach were laid out by Constantin {\it et al.} 
\cite{Con91,ConPro93,ConPro94} who derived the fractal dimension 
$\delta_g^{(d)}$ ($d$ is the space dimension).  Closely related to the present 
investigation is the application to two-dimensional chaotic surface 
waves\cite{ProCon93}.  The $Pr$ dependence of 3-d passive scalar advection 
within this approach was discussed in \cite{GroLoh94}.  As in that work we will 
aim at a rather direct relation between scaling exponents and velocity structure 
functions. 
 
The outline of the paper is as follows.  In Sec.  \ref{sec_basic} the basic 
concepts of the evaluation of the fractal graph dimension are summarized.  The 
results of the mean-field approach\cite{GroMer92} for fully developed 
two-dimensional turbulence in the direct enstrophy cascade range -- the scaling 
behavior of the 2nd order velocity structure function $D_{2}(r)= \langle|{\bf 
u}({\bf x}+{\bf r},t)-{\bf u}({\bf x},t)|^{2}\rangle$ -- are recalled.  In a 
second step we interpolate the scaling of $D_{2}(r)$ to the inverse energy 
cascade range, where no analytical result is known.  We obtain $D_2(r)$ from 
Fourier transform of an energy spectrum as is found in many numerical 
simulations.  In Sec.  \ref{sec_resu} the fractal dimension of the passive 
scalar graph is derived over a broad range of Prandtl numbers, both in the 
enstrophy inertial subrange (ISR) and in the energy ISR with the previous 
relations for the structure function.  We conclude with a summary, a discussion 
of the relation to the findings in the quasi-two-dimensional dispersion 
experiments by Cardoso {\it et al.}\cite{Car96}, and some remarks on open 
questions.

\section{Basic concepts}  
\label{sec_basic}  
\subsection{Fractal dimension of the passive scalar graph}   
%
{}From now on all considerations are made for the case of a two-dimensional 
flow field.  The graph of the scalar field is then a 2-d surface in 3-d space. 
The Hausdorff dimension of this graph is obtained from the scaling behavior of 
the Hausdorff volume $H(G(B_r^{(2)}))$ of the graph $G(B_r^{(2)})=\{({\bf 
x},\theta)|{\bf x}\in B_r^{(2)},\, \theta=\theta({\bf x})\}$ over a disk of 
radius $r$ (the 2-d ball $B_r^{(2)}$) \cite{Fal85}, 
\begin{eqnarray}  
\label{scal}  
H(G(B_r^{(2)}))\sim r^{\delta_g^{(2)}}\;.  
\end{eqnarray}  
In two dimensions the fractal dimension $\delta_g^{(2)}$ is connected to the 
scaling exponent $\zeta^{(\theta)}_1$, cf.  eq.  (\ref{strucdef}), through the 
inequality\cite{ConPro93} 
\begin{equation}  
\label{ineq1}  
\delta_g^{(2)}\le 3-\zeta^{(\theta)}_1\;.  
\end{equation}  
We assume equality in (\ref{ineq1})\cite{ConPro93,GroLoh94}   
and use the relation  
$\delta_g^{(1)}=\delta_g^{(2)}-1$, where $\delta_g^{(1)}$ is the fractal  
dimension of the level sets $\theta_0=\theta({\bf x})$.  The relative  
Hausdorff volume $H(G(B_r^{(2)}))/V(B_r^{(2)})$ is given by geometric measure  
theory\cite{Fed69,Mor88} as  
\begin{eqnarray}  
\label{vol}  
\frac{H(G(B_r^{(2)}))}{V(B_r^{(2)})}&=&  
\frac{1}{V(B_r^{(2)})}\int_{B_r^{(2)}}\,  
\sqrt{1+r^2|{\bf\nabla}\tilde{\theta}|^2}\,\mbox{d}^2{\bf x}\;, \nonumber \\  
&\le&\sqrt{1+\frac{1}{\pi} \int_{B_r^{(2)}}\,|{\bf\nabla}\tilde{\theta}|^2  
\,\mbox{d}^2{\bf x}}\;,  
\end{eqnarray}  
where the Cauchy-Schwartz inequality and $V(B_r^{(2)})=\pi\,r^2$ were used in 
the last line.  The passive scalar field $\theta({\bf x},t)$ is measured in 
units of $\theta_{r.m.s.}=\sqrt{\langle\theta^2\rangle}$, thus leading to 
dimensionless $\tilde{\theta}=\theta/\theta_{r.m.s.}$.  Equation (\ref{vol}) is 
a generalization of the well-known volume formula 
$V=\int\sqrt{g}\,\mbox{d}^2{\bf y}$ to fractal sets, where $V$ is a 
two-dimensional curved hyper surface embedded in the three-dimensional Euclidean 
space and $g$ the determinant of the metric tensor $g_{ij}(y^1,y^2)$. 
 
We now turn to the evaluation of $\delta_g^{(2)}$. The term   
$|{\bf\nabla}\tilde{\theta}|^2$ can be replaced by means of (\ref{sceq}) 
by 
\begin{equation}  
|{\bf\nabla}\tilde{\theta}|^2=\frac{1}{2\kappa}(\kappa  
\Delta\tilde{\theta}^2-({\bf u}\cdot{\bf \nabla})\tilde{\theta}^2)+  
\frac{f_{\theta}\tilde{\theta}}{\kappa\theta_{r.m.s.}}\;.  
\end{equation}  
With this eq.~(\ref{vol}) becomes 
\end{multicols} 
\begin{eqnarray}  
\label{vol1}  
\frac{H(G(B_r^{(2)}))}{V(B_r^{(2)})}  
\le\sqrt{1+\frac{1}{\pi} \int_{B_r^{(2)}}\,  
\left\{\frac{1}{2\kappa}[-({\bf u}\cdot{\bf \nabla})\tilde{\theta}^2  
+\kappa\Delta\tilde{\theta}^2]+  
\frac{f_{\theta}\tilde{\theta}}{\kappa\theta_{r.m.s.}}\right\}  
\,\mbox{d}^2{\bf x}}\;. 
\end{eqnarray}  
\begin{multicols}{2}
We will consider the three integrals under the square root separately  
and denote them by  $I_1$, $I_2$, and 
$I_3$, respectively.  In the three-dimensional case\cite{GroLoh94} the terms 
$I_2$ and $I_3$ vanish in the large Reynolds number limit.  They also satisfy 
the inequality $I_2\le3\sqrt{I_3}$ which changes to $I_2\le2\sqrt{I_3}$ in the 
two-dimensional case.  $I_3$ can be estimated as 
\begin{eqnarray}  
\label{i3eq}  
I_3=\frac{1}{\pi}\int_{B_r^{(2)}}\,\frac{f_{\theta}\tilde{\theta}}  
         {\kappa\theta_{r.m.s.}}\,\mbox{d}^2{\bf x}   
   =\frac{r^2\epsilon_{\theta}}{\kappa\theta_{r.m.s.}^2}  
   =\frac{\epsilon_{\theta}\epsilon_{\omega}^{-1/3}}{\theta_{r.m.s.}^2}  
        Pr\,\tilde{r}^2\;,  
\end{eqnarray}  
where the scalar dissipation rate 
$\epsilon_{\theta}=\kappa\langle|{\bf\nabla}\theta|^2 \rangle$, the enstrophy 
dissipation rate 
$\epsilon_{\omega}=\nu\langle|{\boldmath\nabla}\omega|^2\rangle$, and 
stationarity are used.  In the case of a three-dimensional passive 
scalar this term contains a factor $\nu^{1/2}$ and thus can be neglected. 
In 2-d the smallest scales are given by the enstrophy dissipation rate 
and this factor disappears. Hence $I_3$ cannot be neglected; 
its importance is evidently controlled by Prandtl number $Pr$, 
length scale $r$ and dimensionless prefactor 
\begin{equation}  
\alpha = \frac{\epsilon_{\theta}\epsilon_{\omega}^{-1/3}} 
{\theta_{r.m.s.}^2} \,. 
\label{alp_def} 
\end{equation} 
The term $I_2$ can still be neglected on 
account of its subdominant scaling in $r$.  We introduce dimensionless length 
scales $\tilde{r}=r/\eta_{\omega}$ by means of the enstrophy dissipation length $\eta_{\omega}=\nu^{1/2} 
\epsilon_{\omega}^{-1/6}$ since in 2-d turbulence it is the enstrophy cascade 
that brings the energy to the smallest scales where viscosity dominates. 
 
It follows from (\ref{vol1}) for $I_1$ by applying the Gauss Theorem and the  
Cauchy-Schwarz inequality  
\begin{eqnarray}  
I_1&=& \frac{r}{\kappa}\oint_{\partial B_r^{(2)}}  
                      \frac{\tilde{\theta}^2  
                      ({\bf u}-{\bf u}_0)\cdot{\bf n}}{u_r}\,\mbox{d}\,r\;,  
\nonumber \\  
&\le&\frac{r}{\kappa}  
\sqrt{\oint_{\partial B_r^{(2)}}\frac{\tilde{\theta}^4}{u_r}\,\mbox{d}\,r}   
          \;  
          \sqrt{\oint_{\partial B_r^{(2)}}\frac{  
                                        (({\bf u}-{\bf u}_0)\cdot{\bf n})^2  
           }{u_r}\,\mbox{d}\,r}\;.  
\end{eqnarray}  
The quantity $u_r=2\pi r$ is the circumference.  It is possible to add ${\bf 
u}_0$,  the velocity at the center of $B_r^{(2)}$, due to the assumed 
homogeneity.   
 
The first term on the right hand side contains the square root of the 
passive scalar flatness. Since we are interested in the scaling properties  
of $I_1$, it suffices to know that the scalar flatness is 
a constant, independent of $r$. However, there do not seem to  
be numerical or experimental data for the passive scalar flatness in 2-d. 
Data for the velocity field from the experiments 
\cite{JerTab97} and the numerical simulations \cite{SmiYak93}  
suggest Gaussian behavior in the absence of coherent structures 
in the regime of the inverse cascade.  
More recent experiments suggest that this result also 
extends into the region of the direct enstrophy cascade \cite{Tab99}.  
However, since there are models where a Gaussian statistics for a random 
velocity field causes non-Gaussian  
scalar statistics \cite{Sigg94,Krai94}, this information is insufficient 
to infer Gaussian statistics for the passive scalar. In the following 
we will work with the Gaussian flatness value of three  
for the passive scalar. It should be kept in mind that  
deviations from this value will most 
likely be scale dependent and will give rise to modifications 
of the scaling exponents.  
 
The second term is the longitudinal velocity structure 
function $D_{\parallel}(r)$.  Thus we find 
\begin{equation}  
\label{i1in}  
I_1\le\frac{\sqrt{3}}{\kappa}r\sqrt{D_{\parallel}(r)}\;.  
\end{equation}  
Combining (\ref{scal}), (\ref{vol1}), (\ref{i3eq}), and  
(\ref{i1in}) we end up with an inequality for  
the fractal dimension $\delta^{(2)}_g$ of the passive scalar graph in two  
dimensions, 
\begin{eqnarray}  
\label{fracdim}  
\delta^{(2)}_g-2\le\frac{\mbox{d}}{\mbox{d}\,\ln\tilde{r}}  
\ln\sqrt{1+\frac{\epsilon_{\theta}\epsilon_{\omega}^{-1/3}}{\theta_{r.m.s.}^2}  
        Pr\,\tilde{r}^2+\sqrt{3} Pr \tilde{r}\sqrt{\tilde{D_{\parallel}}}}  
        \;,  
\end{eqnarray}  
where $\tilde{D}_{\parallel}=D_{\parallel}/(\epsilon^{2/3}_{\omega} 
\eta_{\omega}^{2})$.  This inequality, relating the scaling exponent 
$\delta^{(2)}_g$ to the longitudinal structure function of the underlying 
turbulent flow field $\tilde{D_{\parallel}}$ is the main result of this section. 
For most of the discussion that follows we will assume equality in 
(\ref{fracdim}); in the three-dimensional case this is a very good assumption 
\cite{GroLoh94}.

\subsection{Structure functions in two-dimensional turbulence}   
To evaluate (\ref{fracdim}) we need information on the scaling behavior of the 
2nd order longitudinal structure function $D_{\parallel}$. The longitudinal 
structure function $D_{\parallel}(r)$ 
and transversal structure function $D_\perp(r)$ make 
up the velocity structure function $D_2(r)$ and are connected by 
incompressibility, $D_{\perp}=D_{\parallel}+r\frac{dD_{\parallel}}{dr}$. 
Eliminating the transversal part then gives \cite{MonYag75,Lan87} 
\begin{eqnarray}  
\label{dlon}  
D_{\parallel}(r)=\frac{1}{r^2}\int_0^r \rho\,D_2(\rho)\,\mbox{d}\,\rho\;.  
\end{eqnarray}  
  
As there are two 
inertial ranges with several different scaling regimes, there is no analytical 
expression for the structure function.  As far as we are aware, the best that 
can be achieved analytically is the structure function for the enstrophy cascade 
as discussed by Grossmann and Mertens\cite{GroMer92}.  They used a mean field 
type approach for the fully developed, turbulent velocity field in the enstrophy 
cascade, i.e.  for spatial scales $\eta_{\omega}<r<r_{in}$.  Separating small and 
large scales one finds energy and enstrophy balance equations where terms 
resulting from the small scale fluctuations act like an effective eddy viscosity 
for the large scale components of $\omega$.  Analytical expressions for the 2nd 
order vorticity structure function $D^{(\omega)}_2(r)$ and the 2nd order 
velocity structure function $D_2(r)$ can be found using the Batchelor 
interpolation technique\cite{GroMer92,Bat51}.  In dimensionless form they read 
\end{multicols}
\begin{eqnarray}  
\label{mergro1}  
\tilde{D}_2(\tilde{r})&=&\frac{\tilde{D}_2^{(\omega)}(\infty)}{4}\,  
         \frac{\tilde{r}^{2}}{(1+a\,\tilde{r}^{2})^{1/3}}+  
         \left(\frac{Re^{\ast}}{2}-  
         \frac{\tilde{D}_2^{(\omega)}(\infty)}{4}\right)\,\tilde{r}^{2}  
         \;,  
\end{eqnarray}  
\begin{multicols}{2} 
with the parameter $a=\frac{15}{592}$ and the asymptotic value 
$\tilde{D}^{(\omega)}_2(\infty)= 
D^{(\omega)}_2(\infty)/\epsilon_{\omega}^{2/3}=14.8$.  This spectrum also 
depends on the energy dissipation $\epsilon$, which when expressed in the length 
and energy scales of the enstrophy cascade becomes the dimensionless parameter 
$Re^{\ast}=\epsilon/(\epsilon_{\omega}^{2/3}\nu)$.  The structure functions are 
shown in Fig.~1.  Besides the prominent $\tilde{r}^2$ behavior that follows 
already by dimensional analysis one notes an intermediate scaling with 
$\tilde{r}^{4/3}$; the range over which this scaling is observed depends on 
$Re^\ast$ (see below).  The corresponding longitudinal velocity structure 
function $\tilde{D}_{\parallel}(\tilde{r})$ is given with eq.  (\ref{dlon}) by 
\begin{eqnarray}  
\label{mergro2}  
\tilde{D}_{\parallel}(\tilde{r})&=&  
         \left(\frac{Re^{\ast}}{8}-  
         \frac{\tilde{D}_2^{(\omega)}(\infty)}{16}\right)\,\tilde{r}^{2}+  
         \frac{3\tilde{D}_2^{(\omega)}(\infty)}{8 a}\nonumber\\  
                               &\times&\left[  
         \frac{(1+a\,\tilde{r}^{2})^{5/3}-1}{5\,a\,\tilde{r}^{2}}-  
         \frac{(1+a\,\tilde{r}^{2})^{2/3}-1}{2\,a\,\tilde{r}^{2}}  
                                       \right]\;.  
\end{eqnarray}  
  
For the energy ISR no such analytical expression is known.  We therefore combine 
a model for the energy distribution in $k$-space with numerical transformations 
to obtain the longitudinal structure function.  Recent experiments on forced 
two-dimensional turbulence\cite{JerTab97,Rut98}, a number of direct numerical 
simulations\cite{SmiYak93,FriSul84,Ben86,Bor93,KevFar97,Bab97}, field theoretical  
investigations\cite{Fal94} as well as cascade 
models\cite{Schu94} support the existence of a Kolmogorov-like scaling for the 
energy spectrum, $E(k)\sim\,k^{-5/3}$ for $(k<k_f)$, in the energy ISR and 
$E(k)\sim\,k^{-\beta}$ with $\beta\ge3$ for $(k>k_f)$ for the enstrophy ISR.  We 
therefore start with the following model spectrum for the amplitudes 
$\langle|{\bf u}_{\bf k}|^2\rangle$ of the velocity field in a Fourier 
representation in a periodic box of size $L=2\pi$ 
\end{multicols}
\begin{eqnarray} 
\langle|{\bf u}_{\bf k}|^2\rangle\!\sim\! 
\left\{  
\begin{array}{r@{\quad:\quad}l}  
k^3             & \frac{2\pi}{L}\le k\le k_1, \\  
k^{-2/3}        & k_1<k\le k_f,\\  
k^{-\beta}      & k_f<k\le k_{\omega}=\frac{1}{\eta_{\omega}},  
\;\beta\ge 2\, ,\\  
k^{-\beta}\exp\left[-\left(\frac{k-k_{\omega}}{k_{\omega}}\right)^{2}\right]  
                & k_{\omega}<k\;.  
\end{array} 
\right. 
\label{modspec} 
\end{eqnarray} 
\begin{multicols}{2}
Note the different scalings for $\langle|{\bf u}_{\bf k}|^2\rangle$ and the 
energy spectrum $E(k)$ due to phase space factor, i.e.   
$E(k)\!\sim\!k^{-\beta-1}$ 
corresponds to $\langle|{\bf u}_{\bf k}|^2\rangle\!\sim\!k^{-\beta}$. 
 
The first range approximates finite system size effects where we have chosen a 
slope of 3 in correspondence with results of numerical 
experiments\cite{FriSul84,Bab97}.  This is followed by the inverse energy cascade 
range with a Kolmogorov-like scaling law.  At the injection scale $k_f$ the 
enstrophy cascade to larger values of $k$ starts, followed by the viscous 
cutoff.  The energy spectra with $\beta=3$ for three different values of the 
injection wavenumber $k_f$ are shown in Fig.  \ref{p5}. 
 
The relation between velocity spectrum scaling and the velocity structure  
function $D_2(r)$ assuming stationarity, homogeneity, and isotropy is given   
by the volume average  
\begin{eqnarray}  
D_2(r)&=&\frac{1}{V}\int_{V}\,|\,{\bf u}({\bf x}+{\bf r})-{\bf u}({\bf x})\,|  
       ^{2}\,\mbox{d}\!V,\nonumber \\  
      &=&\frac{1}{V}\int_{V}\,|\,\sum_{{\bf k}}{\bf u}_{\bf k}\,  
       \exp(i{\bf k}\cdot{\bf x})  
      [\exp(i{\bf k}\cdot{\bf r})-1]\,|^{2}\,\mbox{d}\!V,\nonumber \\  
      &=&2\,\sum_{{\bf k}}\,\langle|{\bf u}_{\bf k}|^{2}\rangle\,  
       (1-\cos{({\bf k}\cdot{\bf r})})\, .  
\end{eqnarray}    
By averaging over all directions (due to isotropy) in ${\bf k}$-space   
the cosine gives rise to the Bessel function $\mbox{J}_0(kr)$,  
\begin{equation}  
\label{mod_struc}  
D_2(r)=2\,\sum_k\,\langle|{\bf u}_{\bf k}|^{2}\rangle\,  
(1-\mbox{J}_{0}(kr)) \,.  
\end{equation}  
The model spectrum (\ref{modspec}) is then substituted and the summation in 
(\ref{mod_struc}) is evaluated numerically using a finite, geometrically 
scaling set 
of wave numbers.  It should be mentioned here that the model does 
not contain a spectral range 
that would correspond to the intermediate $\tilde{r}^{4/3}$--scaling of 
the structure function in the enstrophy ISR found in the analytical theory. 
We will come back to this point in the discussion of our results.

\section{Results}  
\label{sec_resu}  
\subsection{Fractal dimension in the enstrophy ISR}   
We first calculate the scaling behavior in the enstrophy ISR where the analytical 
expression (\ref{mergro1}) is available.  Inserting (\ref{mergro1}) in 
(\ref{fracdim}) and neglecting the term $I_3$ for the moment, one notes that 
$\delta_g^{(2)}$ depends on three quantities:  the parameter $Re^{\ast}$, the 
Prandtl number $Pr$, and the scale $\tilde{r}$ itself.  The numerical result for 
$\delta_g^{(1)}=\delta_g^{(2)}-1$ are shown in Fig.  \ref{p2} for a Prandtl 
number range varying over ten orders of magnitude and $Re^{\ast}=7.6$.  The grey 
shaded area denotes the range of scales where $\tilde{r}^{4/3}$ gives the main 
contribution to the structure function.  It is only in this range that we find 
$1<\delta_g^{(1)}<2$.  The range is bounded by 
$\tilde{r}_1\le\tilde{r}\le\tilde{r}_2$ where $\tilde{r}_1$ is the crossover 
scale from the viscous subrange (VSR) and $\tilde{r}_2$ is the crossover scale 
to the $\tilde{r}^2$--scaling in the enstrophy ISR, 
\begin{eqnarray}  
\label{scaran}  
\tilde{r}_1 &=&\frac{1}{\sqrt{3}} 
(\tilde{D}_2^{(\omega)}(\infty))^2(Re^{\ast})^{-3/2}\;,  
\nonumber \\  
\tilde{r}_2 &=&\frac{1}{\sqrt{3}} 
(\tilde{D}_2^{(\omega)}(\infty))^2(Re^{\ast}- 
\tilde{D}_2^{(\omega)}(\infty)/2)^{-3/2}\;,  
\end{eqnarray}  
where $\tilde{D}_2^{(\omega)}(\infty)=14.8$ has to be taken.  The larger 
$Re^{\ast}$ the smaller the range of the $\tilde{r}^{4/3}$--scaling.  It can be 
observed only for $Re^{\ast}$ within the interval 
\begin{eqnarray} 
\label{rebounds} 
7.4\!\approx\!\frac{\tilde{D}_2^{(\omega)}(\infty)}{2}  
\le Re^{\ast}\le  
\left(\frac{(\tilde{D}_2^{(\omega)}(\infty))^4}{3}\right)^{1/3}\!\approx\!25\;.  
\end{eqnarray}  
The lower bound follows from the positivity of the structure function by its 
definition (cf.  second term of (\ref{mergro1})).  The upper bound is a result 
of eq.  (\ref{scaran}) and the constraint $\tilde{r}_1\ge1$.  For $Re^{\ast}$ 
approaching 7.4 follows $\tilde{r}_2$ goes to infinity.  The $\tilde{r}^{4/3}$-- 
scaling range is then extended over the whole enstrophy ISR.  We see in Fig. 
\ref{p3} that for increasing $Re^{\ast}$ the intermediate fractal scaling of the 
graph is more and more suppressed and conclude that this behavior of 
$\delta^{(1)}_g$ is due to the presence of the $\tilde{r}^{4/3}$--scaling range. 
The above estimates give $\tilde{r}_1\!\approx\!6.0$ and 
$\tilde{r}_2\!\approx\!1400$ for $Re^{\ast}=7.6$ and 
$\tilde{r}_1\!\approx\!1.0$ and 
$\tilde{r}_2\!\approx\!1.7$ for $Re^{\ast}=25.0$, respectively. 
In the lower panel the corresponding 
scaling exponent of the scalar structure function 
$\zeta^{(\theta)}_1=2-\delta^{(1)}_g$ is plotted.  The plateau of the structure 
function $D_1^{(\theta)}$ for large Prandtl number and scales below the smallest 
scales in the turbulent fluid ($r/\eta_\omega<1$) corresponds to the Batchelor 
regime of chaotic scalar advection in a smooth fluid\cite{Bat59}. 
 
For {\it small values of $Pr$} the diffusion $\kappa$ dominates the passive 
scalar dynamics.  The scalar field is smooth, $\delta^{(1)}_g=1$.  The exponent 
$\delta^{(1)}_g$ grows when the second term in the square root of eq. 
(\ref{fracdim}) becomes dominant.  By inserting the power law 
$\tilde{D}_{\parallel}=\frac{\sqrt[3]{9}}{20} 
(\tilde{D}_2^{(\omega)}(\infty)\tilde{r})^{4/3}$ for the enstrophy ISR at 
$\tilde{r}=\tilde{r}_c$ one gets a crossover for 
\begin{eqnarray}  
\tilde{r}_c=\sqrt[10]{\frac{8000}{243}}Pr^{-3/5}(\tilde{D}_2^{(\omega)} 
(\infty))^{-2/5}  
\approx 0.48\, Pr^{-3/5}. 
\label{rcsmooth}
\end{eqnarray}  
By putting $\tilde{r}_c=\tilde{r}_2$ and using (\ref{scaran})  
the maximum Prandtl number $Pr_s$ without fractal $\delta^{(1)}_g$ can  
be estimated as  
\begin{eqnarray}  
Pr_s<2\sqrt{5}(\tilde{D}_2^{(\omega)}(\infty))^{-4} 
(Re^{\ast}-\tilde{D}_2^{(\omega)}(\infty)/2)^{5/2}\;.  
\end{eqnarray}  
With $Re^{\ast}=7.6$ and $25.0$ this gives   
$Pr_s\lesssim 2\cdot 10^{-6}$ and $10^{-1}$, respectively.  
  
For {\it large values of $Pr$} one observes a transition to $\delta^{(1)}_g=2$ 
even when the velocity field is in the VSR.  Again the second term of 
(\ref{fracdim}) dominates because of its large prefactor $Pr$.  Taking 
$\tilde{D}_{\parallel}=\frac{Re^{\ast}}{8}\tilde{r}^2$ for the VSR gives 
\begin{eqnarray}  
\tilde{r}_c=\sqrt[4]{\frac{8}{3}}Pr^{-1/2}(Re^{\ast})^{-1/4}\;.  
\end{eqnarray}  
With $\tilde{r}_c=\frac{1}{10}\tilde{r}_1$ we get those $Pr_l$ which give 
$\delta^{(1)}_g=2$ in the VSR over at least one decade of scales, 
\begin{eqnarray}  
Pr_l>200\sqrt{6}(\tilde{D}_2^{(\omega)}(\infty))^{-4}(Re^{\ast})^{5/2}\;.  
\end{eqnarray}  
For $Re^{\ast}=7.6$ and $25.0$ this results in $Pr_l\!\gtrsim\!2.0$ and $30.0$, 
respectively. 
 
The structure function of a passive scalar in the enstrophy ISR shows four 
different regimes.  For very small $\tilde{r}$ smoothness gives 
$\delta_g^{(1)}=1$.  This is followed by the Batchelor regime $\delta_g^{(1)}=2$ 
for sufficiently large $Pr$.  The $\tilde{r}^{4/3}$--scaling discovered by 
Grossmann and Mertens is reflected in a decrease of $\delta_g^{(1)}$ below $2$ 
near $r/\eta_\omega\approx 10^{1\pm1}$.  For larger $\tilde{r}$ it goes back up 
to $2$. 
 
So far we neglected the term $I_3=\alpha \,Pr\,\tilde{r}^2$  
(see (\ref{i3eq}) and (\ref{alp_def})) in our calculation.   
Because of its $\tilde{r}^2$-scaling it dominates the structure 
function for large $\tilde{r}$.  
In \cite{ProCon93} this term was assumed to be subdominant.  
Substituting the various  
definitions it can be expressed as a ratio of two rates, 
\begin{eqnarray}  
\label{taylor}  
\alpha=\frac{\kappa \langle|{\bf\nabla}\theta|^2 \rangle}  
                {(\nu \langle|{\bf\nabla}\omega|^2 \rangle)^{1/3}  
                \langle\theta^2 \rangle}  
      =\frac{r_{\theta}}{r_{\omega}}\,.  
\end{eqnarray}  
The rate $r_{\theta}=\epsilon_{\theta}/\theta^2_{r.m.s.}$ is 
a scalar forcing rate.  $r_{\omega}=\epsilon_{\omega}^{1/3}$  
is the strain rate in the enstrophy cascade and characteristic of 
the passive scalar advection by the vortices. 
The case $\alpha>1$ then corresponds to $r_\theta>r_\omega$, 
i.e. fast driving and slow advection. Then the scalar field 
fills space and $\delta_g^{(1)}\sim 2$. In the other case, 
$\alpha<1$, the advection dominates and the structure function 
of the fluid is reflected in that of the scalar. 
It is this latter case that was discussed in \cite{ProCon93}  
for surface waves. 
The size of $\alpha$ is determined by the experimental  
situation and has to be taken from 
measurements.  All quantities that enter (\ref{taylor}) are experimentally 
accessible; note that the enstrophy dissipation rate is related to velocity 
gradients via $\epsilon_{\omega}=-8\langle(\partial_x 
u_x)^3\rangle$\cite{GroMer92}. 
 
Results for different $Pr$ with $\alpha=1$ are shown  
in Fig.~\ref{p4}.  The main effect 
of an increasing $I_3$ is the suppression of the crossover scaling 
and a transition for large $r$.

\subsection{Extension to the energy ISR}   
 
The extension of $D_2(r)$ to the whole range of scales is done with 
(\ref{mod_struc}) and the results for $\delta_g^{(1)}$ are given in 
Fig.~\ref{p6} for three input model spectra (see Fig.  \ref{p5}) which differ by 
the injection wavenumber $k_f$.  The smaller $k_f$ the longer is the enstrophy 
ISR extended which results in a dominant range where $\delta_g^{(1)}=2$.  On the 
other hand, the larger $k_f$ the more dominant the inverse energy cascade range, 
indicated as the grey shaded area in Fig.  \ref{p6}.  The corresponding 
longitudinal velocity structure function $D_{\parallel}(r)$ is superimposed. 
Note that the model spectrum has to be normalized to give 
$\tilde{D}_{\parallel}=\frac{Re^{\ast}}{8}\tilde{r}^2$ in the VSR.  In the 
enstrophy ISR we find $\tilde{D}_{\parallel}(r)\sim \tilde{r}^2$ and in the 
energy ISR $\tilde{D}_{\parallel}(r)\sim \tilde{r}^{2/3}$, leading to 
$\delta_g^{(1)}=2$ and $\delta_g^{(1)}=\frac{5}{3}$, respectively.   
As mentioned, the model spectrum does not show the  
$\tilde{r}^{4/3}$--scaling predicted by \cite{GroMer92}. Therefore, 
if $Re^{\ast}$ is in the range where a $\tilde{r}^{4/3}$--scaling appears 
the $\delta_g^{(1)}$ values for $r\simeq\eta_{\omega}$ have to be replaced by 
the ones in Figs.  \ref{p2}, \ref{p3}, and \ref{p4}. 
For very large values of $r$ we can replace $\mbox{J}_{0}(kr)$ by its asymptotic 
form $\mbox{J}_{0}(kr)\approx\sqrt{\frac{2}{\pi kr}} \cos(kr-\frac{\pi}{4})$ 
resulting in $D_2(r)\approx 2\,\sum_k\,\langle|{\bf u}_{\bf k}|^{2}\rangle$ in 
(\ref{mod_struc}).  The constant asymptotic behavior of the structure function 
corresponds with $\delta_g^{(1)}=\frac{3}{2}$ (cf. (\ref{fracdim})). 
 
The model spectrum contains a free parameter $\beta$ which has no 
agreed upon value. Numerical simulations \cite{Ben86,Bor93,KevFar97,Bab97}  
suggest a range $\beta\in[2,4]$. 
For $\beta=2$ we get 
$\delta_g^{(1)}$ slightly below 2 in the enstrophy ISR which changes clearly to 
$\delta_g^{(1)}=2$ for $\beta>2$ (cf.  Fig.  \ref{p7}).  As expected, the value 
of $\delta_g^{(1)}$ in the energy ISR is insensitive to a $\beta$--variation. 
 
Again we have to discuss the additional influence of the $I_3$ term in 
(\ref{vol1}).  Will inverse cascade effects be suppressed in the large $Pr$ 
number case because of the dominance of  
the $\tilde{r}^2$--scaling at large separations? In order to determine 
the scale  $\tilde{r}_c$ where $I_3\ge I_1$, we use the experimental 
value for the  Kolmogorov 
constant $C_K$ \cite{JerTab97} and 
assume a completely extended inverse cascade with no intermittency corrections. 
Then  
$D_2(r)=4\,C_K\,\epsilon^{2/3} \int^{\infty}_0(1-\mbox{J}_{0}(kr))k^{-5/3}\, 
\mbox{d}\,k$ and $\tilde{D_2}=b_2\tilde{r}^{2/3}$. With  
$C_K$ between 5.5 and 7, we find 
$b_2$ between 31.5 and 40 for the energy ISR and thus finally 
\begin{eqnarray}  
\tilde{r}_{\alpha}\ge\left(\frac{9b_2}{8\alpha^2}\right)^{3/4}  
\approx l\alpha^{-3/2}\;, 
\label{ralength} 
\end{eqnarray}  
where $l$ lies between 14 and 17.  The scale $\tilde{r}_{\alpha}$ is shifted 
towards larger values for decreasing $\alpha$.  A factor $\alpha\!\sim\!1$ can 
suppress the scaling behavior in the energy ISR which was found above completely. 
This fact is illustrated in Fig.  \ref{p8}.  Clearly the asymptotic state for 
$\tilde{r}$ to infinity leads here to $\delta_g^{(1)}$ approaching 2. 
 
\section{DISCUSSION}  
\label{sec_sum}  
Our main findings for a passive scalar in a  
2-d turbulent flow field can be summarized as follows:  
(1) There is a critical scale set by equation (\ref{rcsmooth}) below which the spectrum 
is smooth, $\delta_g^{(1)}=1$, because of diffusion dominance. 
(2) Between this scale and the injection scale $r_{in}$ the scaling 
exponent $\delta_g^{(1)}=2$ in most cases. 
(3) An exception is found for $Re^*$ in the interval set by (\ref{rebounds}), 
where a scaling exponent $\delta_g^{(1)}<2$ is found. The limits of this 
interval are given by (\ref{scaran}) and the deviation from 2 is controlled 
by the parameter $\alpha$, eq. (\ref{taylor}). 
(4) Beyond the injection length and up to a length set by 
equation (\ref{ralength}), 
the scalar field scales with the exponent $\delta_g^{(1)}=5/3$ 
as expected for the energy inertial subrange. 
(5) Above the length scale set by (\ref{ralength}), the exponent again increases to  
$2$.  
What is most surprising is that the scaling derived within 
geometric measure theory depends not only on the scaling 
of the velocity field but also on two additional dimensionless numbers, 
the Reynolds number $Re^*$ which causes the intermediate scaling 
in the enstrophy viscous subrange and on $\alpha$ which suppresses 
the velocity field induced scaling at large separations for  
rapid driving. 
 
At this point input from experiments on two-dimensional turbulence is 
necessary to check and expand the theoretical results. 
Cardoso {\it et al.}  \cite{Car96} measured dispersion in a  
quasi-two-dimensional turbulent flow and compared with results 
for the energy inertial subrange. They observed a velocity  
structure function with scaling $\tilde{r}^0$ and a fractal 
dimension $\delta_g^{(1)}$ between $1.3$ and $1.5$ with an average of 
about $1.4$. Substituting a velocity scaling function  
$\tilde{D}_{\parallel}=C\tilde{r}^0$ in our main equation 
(\ref{fracdim}) gives  
\begin{eqnarray}  
\delta^{(1)}_g\le 1+\frac{\mbox{d}}{\mbox{d}\,\ln\tilde{r}}  
\ln\sqrt{1+5000(\alpha \tilde{r}^2+ \sqrt{3 C} \tilde{r})}  
        \;. 
\end{eqnarray}  
If the quadratic term can be neglected, i.e. if $\tilde{r}$ is small 
enough, the inequality reads $\delta^{(1)}_g\le 3/2$. The experimental 
results are indeed below but close to this limit, so that the  
assumption that the distances are small is probably reasonable. 
For larger separation there is a crossover to $\delta^{(1)}_g\le 2$, 
and it would be interesting to see whether the experimental data 
follow this behavior. For the energy inertial subrange  
[and not too large separations, see (\ref{ralength})], the inequality  
would be $\delta^{(1)}_g\le 5/3$, higher than the one for the 
experimentally observed spectrum.  
 
Further experiments or numerical studies to check the results from geometric 
measure theory, especially the ones for the enstrophy cascade and for the 
dependence on $\alpha$, are clearly needed.  Perhaps it is possible to combine 
the experiments on passive scalar mixing \cite{Car96,Goll97}  
with the set-up for 
extended, stationary inverse and direct cascades \cite{JerTab97,Tab99} in order 
to measure the scaling behavior mentioned in (2). In order to check the 
predictions for the enstrophy cascade in (1) the spatial resolution has to be 
enlarged. Otherwise e.g. the existence of the intermediate  
$\tilde{r}^{4/3}$--scaling of $\tilde{D}_2(\tilde{r})$ cannot be detected. 
We remind the reader that this range is only well established for values of 
$Re^{\ast}$ close to its lower threshold (see Fig. \ref{p1}). 
Its localization with respect to $\tilde{r}$ prevents it from being seen  
in the Fourier spectrum, as already discussed by Grossmann and  
Mertens \cite{GroMer92}. 
 
Another open question which calls for more input from numerical simulations  
and experiments is that of the scalar flatness in 2-d. For a non-Gaussian scalar  
statistics we would expect a scale--dependent flatness $F_{\theta}(\tilde{r})$ 
causing a further scale dependence of the third term in (\ref{fracdim}) and 
thus leading to a modification of the present model. 
 
The problem studied here has also interesting links to magnetohydrodynamics. 
First steps towards using geometric measure theory in this context were 
undertaken by Grauer and Marliani\cite{GraMar95}.  In two dimensions there is a 
direct relation between magnetic field advection and the scalar dynamics studied 
here since the vector potential for the magnetic field has only a $z$-component. 
Consequences of this relation are under investigation. 
 

\begin{narrowtext}  
\begin{figure}  
\begin{center}  
\epsfig{file=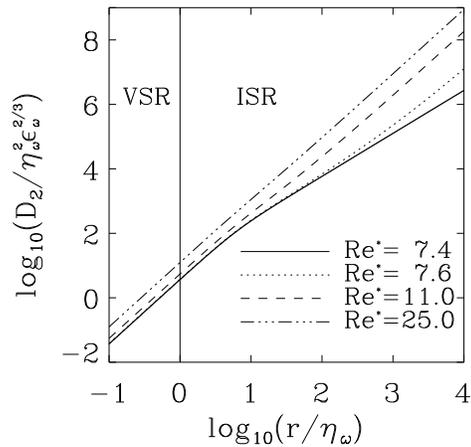 
,width=5cm,height=5cm}  
\end{center}  
\vspace{1cm} 
\caption{Velocity structure function $D_2(r)$ in the enstrophy inertial subrange  
for four different values of $Re^{\ast}$.}  
\label{p1}  
\end{figure}  
\begin{figure}  
\begin{center}  
\epsfig{file=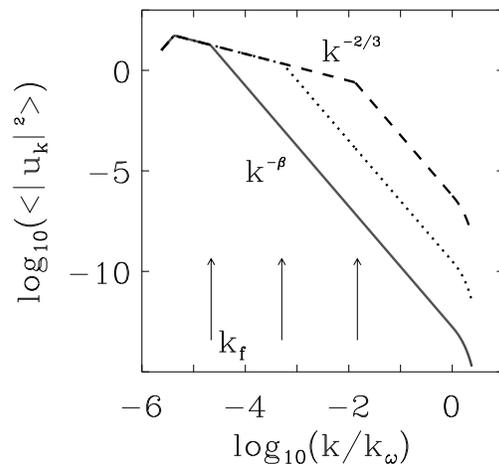,  
width=6cm,height=6cm}  
\end{center}  
\vspace{1cm} 
\caption{Model spectrum $\langle|{\bf u}_{\bf k}|^{2}\rangle$ for three  
different values of $\tilde{k}_f$ indicated by the arrows ($\tilde{k}_f\approx   
2\cdot 10^{-5}, 5\cdot 10^{-4}, 10^{-2}$).  
The wavenumbers are given in units of $k_{\omega}=\eta_{\omega}^{-1}.$ 
The exponent $\beta$ was set to three.}  
\label{p5}  
\end{figure}  
\begin{figure}  
\begin{center}  
\epsfig{file=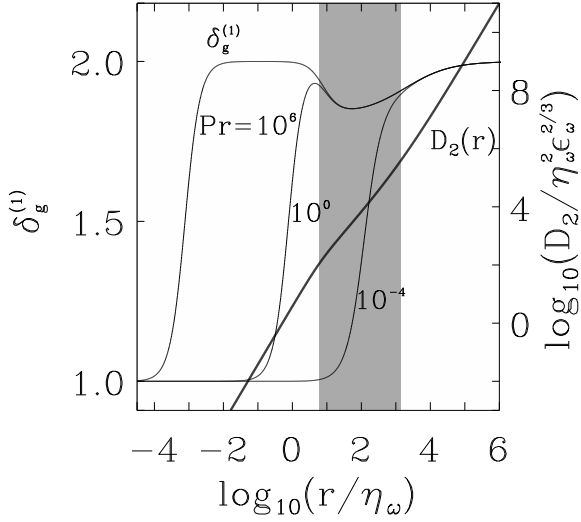 
,width=6cm,height=6cm}  
\end{center}  
\vspace{1cm} 
\caption{Fractal dimension $\delta_g^{(1)}$ for three Prandtl numbers and the  
corresponding velocity structure function $\tilde{D_2}(\tilde{r})$ (thick line)  
for $Re^{\ast}=7.6$.  The grey shaded area denotes the range of scales where the  
$\tilde{r}^{4/3}$ term dominates for the parameter set.  A fractal  
$\delta_g^{(1)}$ can be observed in this range of scales.}  
\label{p2}  
\end{figure}  
\begin{figure}  
\begin{center}  
\epsfig{file=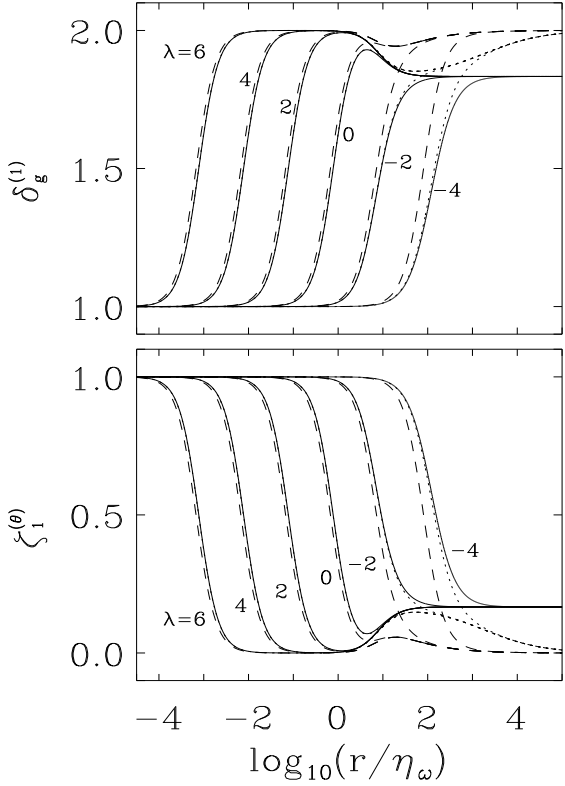, 
width=6cm,height=10cm}  
\end{center}  
\vspace{1cm} 
\caption{Fractal dimension $\delta_g^{(1)}$ and scaling exponent  
$\zeta^{(\theta)}_1$ as a function of $Pr=10^{\lambda}$ and of $Re^{\ast}$. The  
solid line is $Re^{\ast}=7.4$ (the lower bound), the dotted line   
is $Re^{\ast}=7.6$,  
and the dashed line is $Re^{\ast}=13.0$ in both panels. Note that  
$\zeta^{(\theta)}_1=2-\delta_g^{(1)}$.}  
\label{p3}  
\end{figure}  
\begin{figure}  
\begin{center}  
\epsfig{file=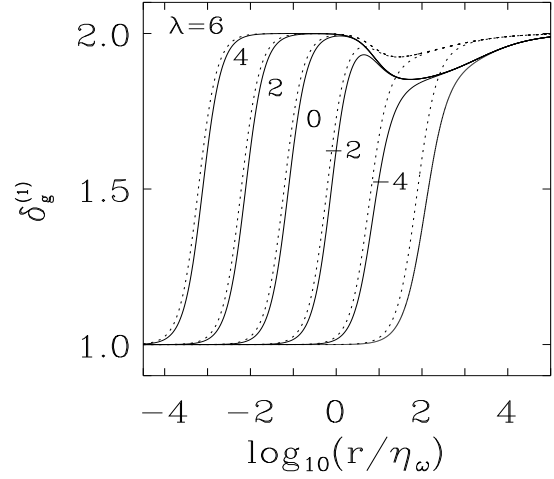, 
width=6cm,height=5.5cm}  
\end{center}  
\vspace{1cm} 
\caption{Fractal dimension $\delta_g^{(1)}$   
as a function of $Pr=10^{\lambda}$ for $Re^{\ast}=7.6$, $\alpha=1$. The  
solid line plots show the results when only the advection term $I_1$   
is taken. The dotted lines show the additional influence of the forcing term  
$I_3$.}  
\label{p4}  
\end{figure}  
\begin{figure}  
\begin{center}  
\epsfig{file=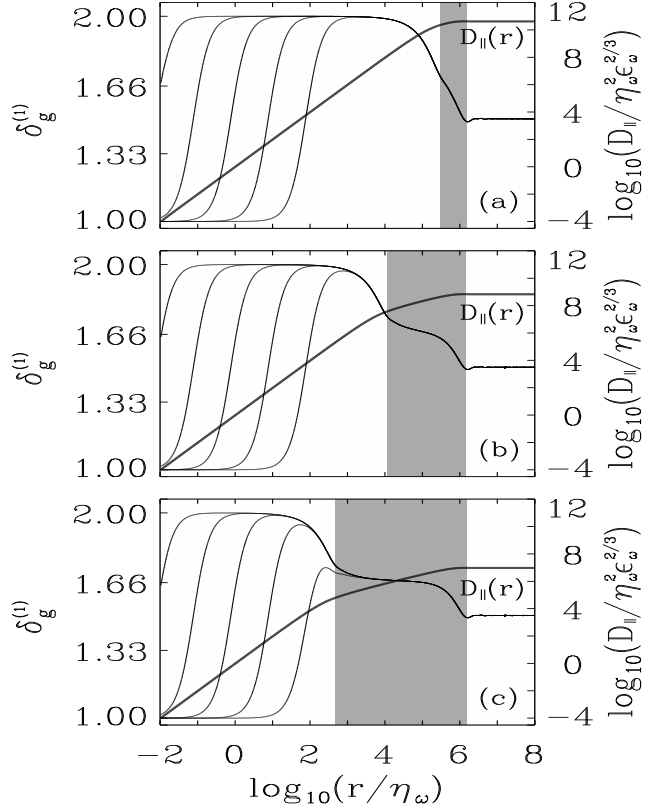,  
width=5.5cm,height=10cm}  
\end{center}  
\vspace{1cm} 
\caption{Longitudinal velocity structure function   
$\tilde{D}_{\parallel}(\tilde{r})$ (thick line) and fractal dimension   
$\delta^{(1)}_g$ over $\tilde{r}$ for $Re^{\ast}=7.6$ and   
$Pr=10^4, 10^2, 10^0, 10^{-2}$, and $10^{-4}$ decreasing from left to right.  
The grey shaded range of scales denotes the inverse  
energy cascade range of $\langle|{\bf u}_{\bf k}|^{2}\rangle$.  
(a): $\tilde{k}_f\approx 2\cdot 10^{-5}$,   
(b): $\tilde{k}_f\approx 5\cdot 10^{-4}$,   
(c): $\tilde{k}_f\approx 10^{-2}$. The exponent $\beta=3$ was taken.}  
\label{p6}  
\end{figure}  
\begin{figure}  
\begin{center}  
\epsfig{file=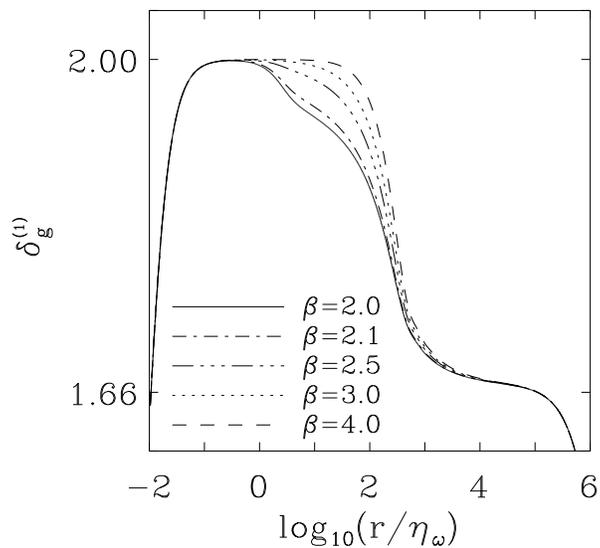,  
width=6.5cm,height=6.5cm}  
\end{center}  
\vspace{1.5cm} 
\caption{Fractal dimension $\delta_g^{(1)}$ for $Pr=10^4$ and   
$Re^{\ast}=7.6$ for different values of the scaling exponent $\beta$  
taken in the enstrophy ISR for the model spectrum   
$\langle|{\bf u}_{\bf k}|^{2}\rangle$ (cf. (\ref{modspec})).}  
\label{p7}  
\end{figure}  
\begin{figure}  
\begin{center}  
\epsfig{file=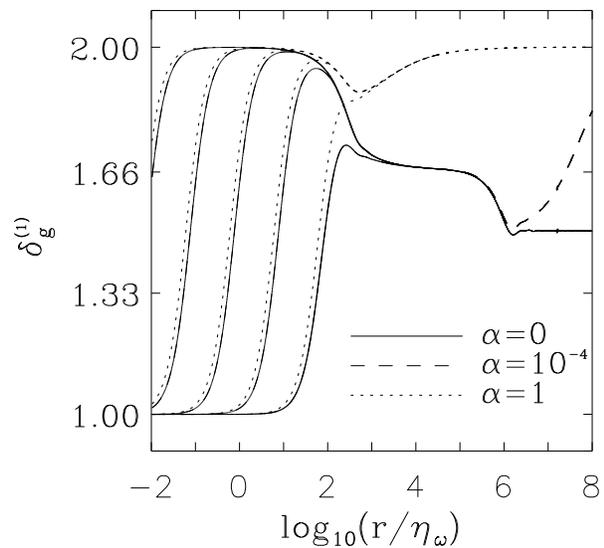,  
width=6.5cm,height=6.5cm}  
\end{center} 
\vspace{1.5cm} 
\caption{Fractal dimension $\delta_g^{(1)}$ for    
$Re^{\ast}=7.6$ and for three different values of the parameter  
$\alpha=\epsilon_{\theta}\epsilon_{\omega}^{-1/3}/\theta_{r.m.s.}^2$.  
The exponent $\beta=3$ was taken.} 
\label{p8}  
\end{figure}  
\end{narrowtext}  
\end{multicols}
\end{document}